\documentstyle[12pt,aasms4]{article}
\newcommand{\etal}{{\it et al.} }
\newcommand{\asca}{{\it ASCA} }

\newcommand{\esrc}{1E~0657$-$56 }
\newcommand{\pks}{1E~0657$-$56 }
\newcommand{\abun}{$A/A_{\odot}$ }

\begin{document}

\title{ON THE UNUSUALLY HIGH TEMPERATURE OF THE CLUSTER OF
 GALAXIES 1E 0657$-$56}

\author{Tahir Yaqoob\altaffilmark{1}}

\vspace{3cm}

\altaffiltext{1}{Laboratory for High Energy Astrophysics, 
NASA/Goddard Space Flight Center, Greenbelt, MD 20771, USA.}

\begin{abstract}
A recent X-ray observation of the cluster \esrc ($z=0.296$) with \asca 
implied an unusually high temperature of $\sim 17$ keV. Such a high
temperature would make it the hottest known cluster and severely
constrain cosmological models since, in a Universe with critical density
($\Omega=1$) the probability of observing such a cluster is 
only $\sim 4 \times 10^{-5}$.
Here we test the robustness of this observational result
since it has such important implications. We analysed the data using 
a variety of different data analysis methods and spectral analysis 
assumptions 
and find a temperature of $\sim 11-12$ keV in all
cases, except for one class of spectral fits. 
These are fits in which the absorbing column density is fixed
at the Galactic value. Using simulated data for a 12 keV cluster,
we show that a high temperature of $\sim 17$ keV is 
artificially obtained if the true spectrum has a stronger low-energy
cut-off than that for Galactic absorption only. 
The apparent extra absorption may be astrophysical in origin,
(either intrinsic or line-of-sight), or it may be a problem with
the low-energy CCD efficiency.
Although
significantly lower than previous measurements, this temperature 
of $kT \sim 11-12$ keV is 
still relatively high since only a few clusters have 
been found to have temperatures higher than 10 keV and the data 
therefore still present some difficulty for an $\Omega=1$ Universe.
Our results
will also be useful to anyone who wants to estimate the systematic
errors involved in
different methods of background subtraction of \asca data for sources
with similar signal-to-noise to that of the \esrc data reported here.
\end{abstract}

\section{INTRODUCTION}

The recently discovered cluster of galaxies, \esrc ($z=0.296$), was 
observed by
{\it ASCA}
({\it Advanced Satellite for Cosmology and Astrophysics};
see Tanaka, Inoue \& Holt, 1994) and found to have an unusually high
temperature of $17.4 \pm 2.5$ keV  (Tucker \etal 1998; hereafter T98).
A full account of this result and other properties of \esrc can be found
in T98. Cosmologies in which the density of the Universe is critical
($\Omega=1$) predict such a small number of clusters with this high a
temperature that an $\Omega=1$ Universe is strongly disfavored
by this single measurement of the temperature of \esrc 
(see T98 and references therein).
Since the cosmological implications are so strong, it is important to
test the robustness of this observational result. So far there are only
two other clusters with temperatures greater than 10 keV 
(MS 1054$-$0321, Donahue \etal 1998; A2163, Markevitch \etal 1996;
both have $kT \sim 12 \pm \sim 2$ keV). Even for the latter objects the
probability of observing $kT>10$ keV for just one source
in an $\Omega=1$ Universe
is less than 0.01 (as opposed to $\sim 4 \times 10^{-6}$ for
\esrc).

In this {\it Letter} we re-analyse the \asca data for \esrc using a
variety of different data analysis methods
and spectral analysis assumptions. The 
different methods of background-subtraction 
and spectral extraction are described in detail in
\S 2 and the spectral analysis of these data is described in \S 3.
Our conclusions are presented in \S 4. 

\section{\asca DATA REDUCTION}

\asca observed \pks in 1996 May 10--11 for a duration
of $\sim 65$ ks. 
Since it is important that our results be
reproducible, 
we describe the data reduction and analysis in some considerable 
detail.   
The reader is referred to Tanaka \etal (1994)
for details of the
instrumentation aboard {\it ASCA}. The two Solid State
Imaging Spectrometers (SIS),
hereafter SIS0 and SIS1, with
a bandpass of $\sim 0.5-10$ keV, were operated in 2-CCD FAINT and BRIGHT
modes. The two Gas Imaging Scintillators (GIS),
hereafter GIS2 and GIS3, with a bandpass
of $\sim 0.7-10$ keV, were operated in
standard PH mode. 
The SIS FAINT and BRIGHT mode data were combined and
the SIS energy scale was fixed using a prescription based on
measurements of Cas A (Dotani \etal 1997). This corrects for 
the continuing decline in the CTI of the CCDs with time
and is based on interpolation or extrapolation of the Cas A 
measurements. The last Cas A measurements incorporated into
the analysis software were made about 
three months after the \pks observation and since the CTI changes
very slowly with time, the interpolation of these
measurements results in a systematic
uncertainty in the gain
which is much less than $1\%$ at 1 keV (Dotani \etal 1997). 
Version 1.1 of the SIS response matrix generator was used in the
analysis. 
However,
no corrections were made for any possible offsets,
fluctuations or distortion of the CCD dark
level distribution (the latter also known as the `RDD' effect).
Such corrections (which themselves are subject to uncertainty) 
can only be applied to FAINT mode data.
This can be done with the software tool `CORRECTRDD' which, however,
as of the time of writing, still does {\it not} correct for the loss in
CCD efficiency. We note that part of our
objective is to reproduce the results of T98 who did not use
CORRECTRDD either because a working version of it was not available
at the time they did their analysis. 
In 1-CCD mode (the SIS observation mode for most point sources) the
RDD effect in SIS0 is negligible for data taken as late as early 1998.
In addition to RDD there are two other effects which lead to a loss
of low-energy efficiency which {\it are} important even in 1-CCD mode.
One of these is due to on-board software mis-grading events
for some settings of
the low-energy discriminator, and the origin of the 
other effect is not yet understood. The latter can reduce the SIS 
efficiency
at 0.6 keV by as much as $\sim 50\%$.
Documentation of these
effects can be found on the World-Wide Web pages
of the \asca Guest Observer Facility (hereafter \asca GOF).
This work is still very much on-going.
For discussions of some calibration issues with respect
to earlier data (but still relevant here)
see Grandi \etal (1997) and Orr \etal (1998).
Considering the finite energy resolution and signal-to-noise of
SIS data, the loss of low energy efficiency (which seems to be
worse in SIS1) can manifest itself as
an apparently
larger measured column density (up to $\sim \rm few \times
10^{20} \rm \ cm^{-2}$). In 2-CCD mode, the 
mode relevant for this observation of 1E 0657$-$56, the effect is 
worse in both SIS0 and SIS1. Typically, for data taken in
late 1997 to early 1998 column densities measured 
for clusters by
SIS0 and SIS1 can be a factor $\sim 2$ and $\sim 4$ 
larger than the corresponding column densities 
measured by the ROSAT {\it PSPC}.
However, a systematic study and development of quantitative
corrections is still underway (see \asca GOF Web pages).

Data were screened so that accepted events satisfied the following criteria:
(i) data were taken outside the SAA; (ii) the time since or before passage
through the SAA or a satellite day/night transition was $>50$ s; (iii)
the elevation angle to Earth was $>5^{\circ}$; (iv) the magnetic 
cut-off rigidity (COR) was $>7\ {\rm GeV/c}$; the deviation of the satellite from
the nominal pointing position was $<0.01^{\circ}$; (v) the SIS parameters
measuring active CCD pixels registered $<100$ active pixels per second,
and (vi) the radiation-belt monitor registered $<500$ ct/s.
Hot and flickering pixels in the SIS
were removed.
Screening resulted in net `good' exposure times in the range $\sim 
23.6$--25.1 Ks for the four instruments.

Source events were extracted from the GIS
using circular regions with radii 6'. For the SIS, two different
types of extraction region were used: (i) circular, with radii
4' (which we call 'C') and (ii) rectangular (which we call 'R'), 
using the exact
dimensions used by T98. The reason for doing this is that no
azimuthal dependence is included in the current XRT responses
so it is useful to check whether there are any systematic errors
introduced by using non-circular extraction regions. 

Background spectra were accumulated in three different ways. The
first, which we call `B4', used 
released blank sky fields which were taken in 4-CCD mode.
Events from the entire field-of-view
were used, excluding the outer ring in the GIS. The second
method (which we call `LS', for
local, same chip) used local background events from the same
chip as the source at radii greater than 5.5' from the source centroid.
For the GIS, annulii centered on the source and having inner
and outer radii
of 8' and 12' respectively were used. The third method (which we
call `LO' for local, off-chip) used the entire off-source SIS chip for
background (slight differences in the spectral responses of the
different chips are negligible compared to statistical and
background-subtraction errors). For the GIS, elliptical regions
with minor and major axes of 6' and 12' respectively, offset 
as far as possible from the source centroid , were used.
All  background data were subject to identical selection criteria
as the on-source data.
Using these different methods of background subtraction will then
test the robustness of any spectral results to differences in the
background in different CCD modes and SIS chips, positions on the sky, epoch,
and contamination from the source itself (in the LS backgrounds).
The background-subtracted spectra extracted from circular regions
resulted in count rates in the range 0.21-0.32 ct/s for
the four instruments, whilst the
smaller, rectangular SIS regions yielded only 0.21 ct/s for SIS0
and 0.14 ct/s for SIS1.  

For the X-ray Telescope (XRT) effective areas we used version 
2.72 of ASCAARF with both empirical effective area factors activated
(these force residuals from fitting \asca data for the Crab Nebula
to be less than 3\%). The `extended source' option was used in 
generating the XRT effective areas. We also tried some preliminary
spectral fitting with responses made using the `point source' option
and found negligible difference in the resulting 
source spectral parameters. From this we conclude that systematics
in the XRT responses due to the spatial extent of \esrc can be neglected 
in comparison to other sources of systematic error.

\section{SPECTRAL FITTING RESULTS}

We fitted spectra from the four \asca instruments simultaneously
in the range 0.5--10 keV with a Raymond-Smith Plasma model, plus 
a cold, neutral absorber at $z=0$. 
A total of three free, interesting parameters were involved (the
plasma temperature, $kT$, plasma abundances relative to solar, $A/A_{\odot}$
(Fe/H = $4.68 \times 10^{-5}$),
and the column density, $N_{H}$. 
The four independent instrument normalizations 
were also free but the deviation
of any of the four normalizations from their mean was less than 20\%.
Such a large systematic difference between fluxes measured by the
SIS and GIS is common for extended sources with photon-limited spectra.
Hereafter we quote mean fluxes and luminosities from the two SIS,
which agree with each other to better than 1\%, and are known to be
more reliable than GIS normalizations.
Note that the Galactic value of the column density, obtained from
Dickey and Lockman (1990) is $6.55 \times 10^{20} \ \rm cm^{-2}$.

First we tested the effect of using the three different
methods of background subtraction (see \S 2) for circular
source extraction regions. The results are shown in Table 1
(fits 1--3). The ratios of the data to the best-fitting model for
fit \#3 is shown in Figure 1. For all three fits we obtain
$kT \sim 11-12$ keV, $N_{H} \sim 15 \times 10^{20} \ \rm cm^{-2}$,
and \abun $\sim 0.20$. For all three parameters, the 
systematic differences arising from the different methods of
background subtraction are less than the statistical errors 
(see Table 1). From fit \#1 we obtain a 0.5--10 keV observed flux  
and intrinsic luminosity of $1.3 \times 10^{-11} \rm \ erg \ cm^{-2}
\ s^{-1}$ and $5.6 \times 10^{45} \ \rm erg \ s^{-1}$ respectively
($H_{0} = 50 \rm \ km \ s^{-1} \ Mpc^{-1}$ and $q_{0} = 0$).
Similar values were obtained for the remaining spectral fits.

Next we tested the effect of using rectangular
extraction regions for the SIS spectra (as in T98), using
the exact dimensions as used by T98 (fit \#4) and obtained
very similar parameters to fits 1--3, except for a slightly lower
column density. Note that the background and region extractions
for fit \#4 are the closest in similarity to those used by T98. 
Thus neither the different background spectra nor different extraction
regions can account for the significantly lower temperatures that
we obtain. Note that our
90\% confidence regions do not overlap with the T98 measurements of  
$17.4 \pm 2.5$ keV.

We then investigated the effect of fixing $N_{H}$ at the Galactic
value. The results for fit \#5 (see Table 1) pertain to the use
of the `LS' background (\S 2) and circular source extraction.
The background (`B4') and extraction regions (`R') 
for fit \#6 are similar to those
used by T98. It can be seen that
the resulting values of $kT$ are now in agreement with T98.
The ratios of data to best-fitting model for fit \#5 are shown in Figure 2
in which the deficit at low energies is clearly seen in the data
(the fit is only marginally acceptable at the 90\% confidence level).
This deficit is not seen in the data presented by T98 because they
used smaller spectral extraction regions (as in fit \#6)
so their statistical errors
were larger. 
Note that T98 do not explicitly state whether the absorption was
fixed at the Galactic value, nor do they give the actual value used,
if it was.

We can easily test whether the high temperature of $\sim 17$ keV
is an artifact of fixing the column density at the Galactic value when
the actual data clearly require a stronger low-energy
turnover (Figure 2). We simulated a 40 ks exposure SIS0 spectrum using an input
model consisting of a Raymond-Smith Plasma with $kT=12$ keV, 
$N_{H} = 14 \times 10^{20} \ \rm cm^{-2}$, and the same flux as
the real \esrc data, and abundances of 0.2 relative to solar. 
We then fitted the simulated data with the same model but with the
column density fixed at the Galactic value and all other parameters
free. We obtained a best-fitting temperature of $17.6^{+3.0}_{-2.6}$ keV.
This clearly demonstrates that if the column density in fitting the
real \esrc data is fixed when the data really have a 
stronger low-energy cut-off
(whether or not this is due to extra intrinsic absorption or a
calibration problem is irrelevant) the result is an artificially high
temperature. Further proof of this comes from 
excluding the low-energy part of the real
data for spectral fitting.
We performed four-instrument fits
in the restricted energy band 1.0--10 keV, with only
Galactic absorption included. This significantly reduces
the impact of any low-energy turnover. We used
background spectra (`B4') and extraction regions (`R')
most closely resembling those of T98 and the results are shown
in Table 1, fit \#7. We obtain 
a lower temperature of $kT=13.8^{+2.2}_{-1.9}$ keV.
We found that excluding even more of the low-energy data
reduced the measured temperature even further.

The next spectral fit in Table 1 (fit \#8) uses SIS spectra
made from only single-pixel CCD events (grade 0) in order to
investigate the effect on the measured temperature
of the less uncertain
calibration of two-pixel events (under some circumstances -
see the \asca GOF Web pages). Table 1 shows that the best-fitting
temperature is still $\sim 12$ keV.

The basic \asca event data used in T98 were produced from the
raw data using `Revision 1' (REV1) of the processing pipeline software.
The event data we used was produced later,
using `Revision 2'
(REV2) of that software (REV1 was phased out). The final spectral fit in
Table 1 (fit \#9) was performed to check whether the 
difference between REV1 and REV2 data could be responsible
for the unusually high cluster
temperature measured by T98.
In practice
the only differences between REV1 and REV2
are minor updates in the calibration files 
for both SIS and GIS that are used to compute temporal gain corrections.
For the SIS, REV2 uses data from new calibration observations
not available at the time REV1 was used (as opposed to extrapolation
using old data). For the GIS, REV2 incorporates
temporal gain corrections not applied in REV1. Another change
in REV2 data is that pre-screened events files are made using
different screening criteria than those used for REV1. However,
this latter difference is not relevant to us because we dot use
pre-screened event data and nor did T98. Fit \#9 in Table 1 is
essentially a repeat of fit \#1 except that fit \#9 uses REV1 data
and fit \#1 uses REV2 data. It can be seen that the resulting
spectral parameters for REV1 and REV2 are indistinguishable. 

\section{SUMMARY AND CONCLUSIONS}

We have re-analysed the \asca data for the cluster of galaxies
\esrc using a variety of different data analysis and spectral fitting
assumptions. The actual methods of data extraction and
background subtraction cover the entire range of methods used, plus more,
in the majority of the
hundreds of papers published using \asca data. The one method of extraction
and analysis
used by Tucker \etal (1998) yielded the unusually high temperature
of $kT = 17.4 \pm 2.5$ keV. All but one of our methods of analysis
gave a much lower temperature, $\sim 11-12$ keV. The only way we
could reproduce the higher temperature was to fix the column density
at the Galactic value, even 
though this leads to a low-energy deficit in the data, which clearly
require a stronger low-energy cut-off. We proved that fixing the
column density in this way leads to an artificially high temperature
by simulating \asca data for a cluster with $kT=12$ keV and
a column density higher than Galactic, and then re-fitting the data
with the column fixed at the Galactic value. The additional low-energy
cut-off in the real \esrc data may be astrophysical in origin
(intrinsic or line-of-sight absorption) or it may be due to 
as yet uncalibrated loss of CCD efficiency in the SIS. 
The latter is a very difficult
problem because there are a number of time-dependent effects which
are causing a loss in the low-energy CCD efficiency and these are
under on-going investigation by the \asca GOF and \asca instrument
teams. 

Even aside from calibration problems, it is recommended that the
absorbing column density is not fixed at the Galactic value
when spectral fitting \asca data.
This is because doing so assumes there is no other photo-electric
absorption, either along the line-of-sight, or intrinsic. 
If this assumption is wrong then, as we have shown, one obtains
artificially high cluster temperatures. Besides, such a spectral 
fitting assumption 
allows no room for systematic errors in the actual value for
Galactic absorption. Even if there is no other absorption and
no calibration problem, an error in the Galactic column density
will lead to an error in the measured temperature.

We note that our results will also be useful to
anyone who wants to estimate the systematic effects
of different background-subtraction methods for \asca data 
for any source with 
a similar count-rate to that of \esrc during the observation reported
here ($\sim 0.21-0.32$ ct/s). 

The temperature of $\sim 11-12$ keV for \esrc still puts it
amongst the handful of clusters with $kT>10$ keV, along with
MS 1054$-$0321 (Donahue \etal 1998) and A2163 (Markevitch 1996).
As such, their existence still poses a problem for a Universe with
$\Omega =1$ in that the probability of observing only one of these
clusters is $<0.01$.

The author thanks T. Dotani, Richard Mushotzky and Una Hwang for 
valuable discussions, the anonymous referee for some extremely
helpful comments, and
the \asca mission operations team at ISAS, Japan,
and all the instrument teams 
for their dedication and hard work in making these \asca observations 
possible. This research made use of the HEASARC archives at the
Laboratory for High Energy Astrophysics, NASA/GSFC.

\newpage

\begin{deluxetable}{lcccccc}
\tablecaption{Raymond-Smith Plasma Spectral Fits to \esrc}
\tablecolumns{7}
\tablewidth{0pt}
\tablehead{
\colhead{Fit \#} & \colhead{BGD$^{a}$} & \colhead{Region$^{b}$} &
\colhead{$N_{H} \ (10^{20} \rm \ cm^{-2})$} & $kT$(keV) & 
$A/A_{\odot}^{c}$ &  \colhead{$\chi^{2}$/d.o.f.$^{d}$}}

\startdata

1 & LS &  C & $14.9^{+3.5}_{-2.5}$ & $11.7^{+2.2}_{-1.4}$ & 
	$0.20^{+0.14}_{-0.13}$ & 726.4/769 \nl
2 & LO &  C & $15.2^{+2.6}_{-2.3}$ & $11.4^{+2.0}_{-1.3}$ & 
	$0.20^{+0.13}_{-0.13}$ & 745.0/769 \nl
3 & B4 &  C & $15.1^{+2.3}_{-2.2}$ & $11.0^{+1.5}_{-1.3}$ & 
	$0.21^{+0.13}_{-0.11}$ & 748.4/769 \nl
4 & B4 &  R & $13.3^{+2.7}_{-2.7}$ & $11.4^{+2.1}_{-1.5}$ & 
	$0.24^{+0.14}_{-0.15}$ & 637.9/689 \nl
5 & LS & C & 6.55 fixed & $17.6^{+2.2}_{-2.2}$ & 
	$0.20^{+0.18}_{-0.18}$  & 814.1/769 \nl
6 & B4 & R & 6.55 fixed & $16.3^{+2.4}_{-2.1}$ & 
	$0.22^{+0.18}_{-0.20}$ & 670.2/690 \nl
7 & B4 (1-10 keV) & R & 6.55 fixed & $13.8^{+2.2}_{-1.9}$ &
	$0.26^{+0.23}_{-0.18}$ & 577.0/649 \nl
8 & LS (G0) & C & $11.8^{+2.8}_{-2.7}$ &  $11.5^{+2.4}_{-1.6}$ & 
$0.16^{+0.15}_{-0.16}$ & 708.8/703 \nl
9 & LS (REV1) & C & $13.5^{+2.5}_{-2.5}$ &  $11.7^{+2.2}_{-1.4}$ & 
      $0.22^{+0.13}_{-0.14}$ & 714.5/764 \nl

\tablecomments{Errors are 90\% confidence, corresponding to
$\Delta \chi^{2} = 6.251$ or $\Delta \chi^{2} = 4.605$, for three
or two free parameters respectively (apart from normalizations). \nl
$^{a}$ LS = Local background (partial annulus)
from same SIS chip as source; LO = local background from off-source 
SIS chip; B4 = standard 4-CCD mode, released, blank-sky background;
G0 = same as fit (1) but using grade 0 data only (see text for details);
`Rev 1' = same as fit (1) but using  
`Revision 1' processed ASCA data as opposed to `Revision 2'i
(see text for details). \nl
$^{b}$ C = circular source extraction region; R = rectangular source
extraction region with dimensions used by T98. \nl
$^{c}$ Abundances relative to solar (Fe/H = $4.68 \times 10^{-5}$). \nl
$^{d}$ d.o.f. = degrees of freedom. } 
\enddata
\end{deluxetable}

\newpage
\section*{Figure Captions}

\par\noindent
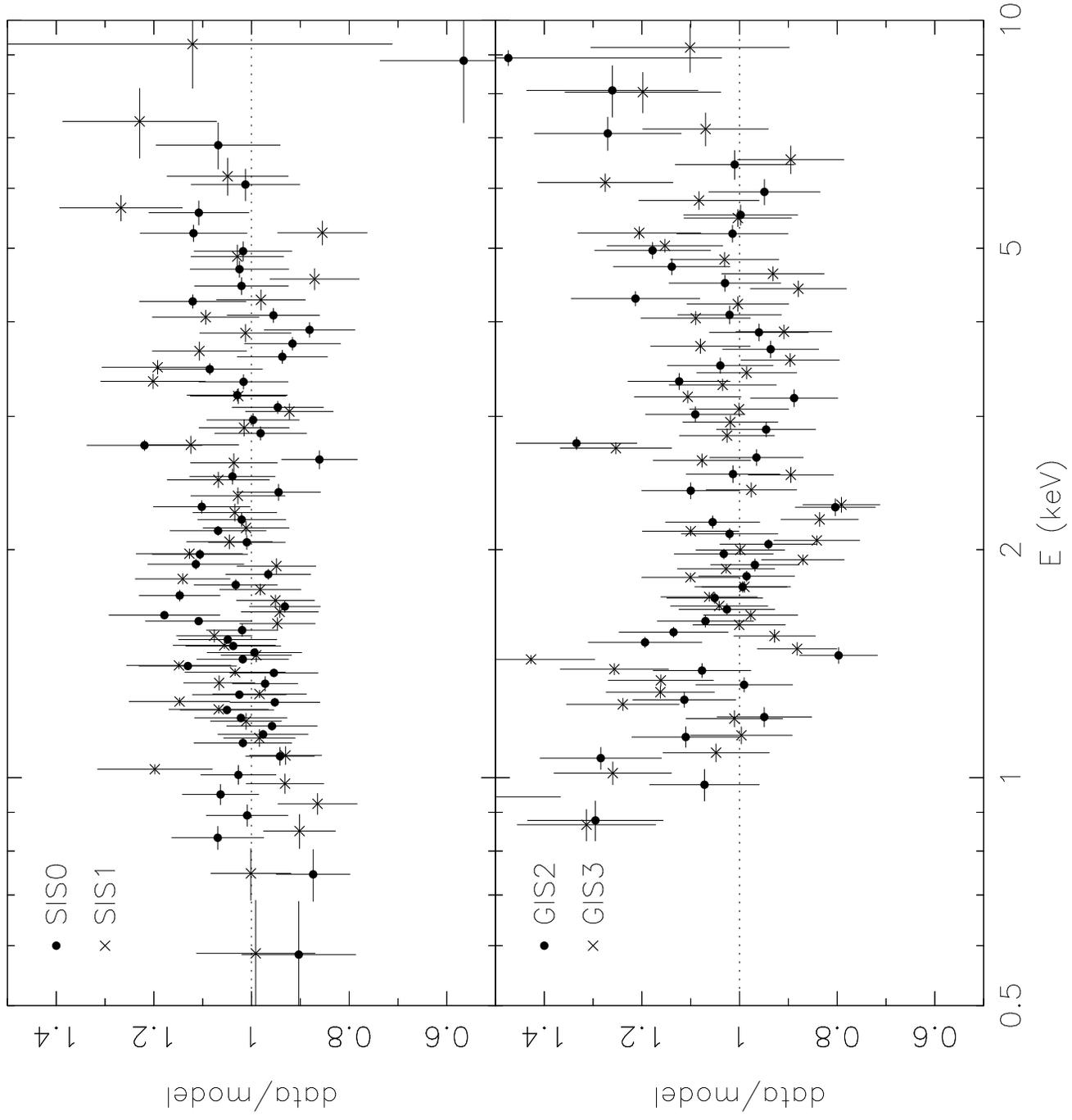
{\bf Figure 1} \\
Ratio of the \asca data for \pks ($z=0.296$)
to the best-fitting Raymond-Smith model (corresponding to fit \#3
in Table 1).  The best-fitting temperature in this case is
$kT=11.0^{+1.5}_{-1.3}$ keV. 

\par\noindent
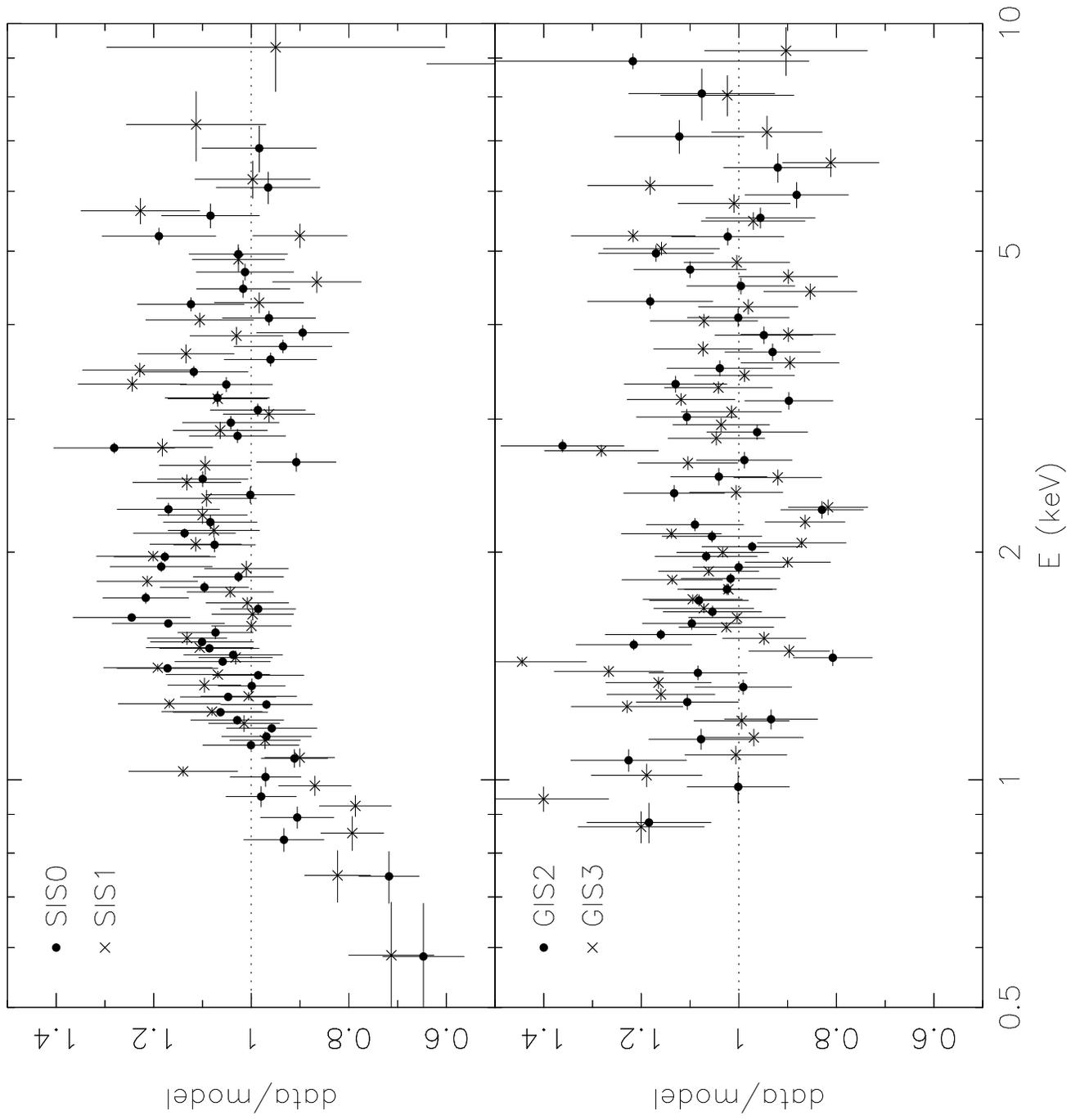
{\bf Figure 2} \\
The result of fitting a Raymond-Smith model to the \pks \asca data
with absorption fixed at the Galactic value (see text).
Shown are the ratios of the \asca data
to the best-fitting Raymond-Smith model (corresponding to fit \#5
in Table 1). The best-fitting temperature in this case is
$kT=17.6^{+2.2}_{-2.2}$ keV.
                                                                               

\begin{references}

\reference{dick1990} Dickey, J.M., \& Lockman, F.J. 
		1990, Ann. Rev. Ast. Astr. 28, 215
\reference{Don98} Donahue, M., Voit, M., Gioa, I., Luppino, G., 
Hughes, J. P., \& Stocke, J. T. 1998, ApJ, 502, 550
\reference{Do97} Dotani, T., \etal 
		1997, {\it ASCANews}, 5, 14
\reference{Paula97} Grandi, P., \etal 
		1997, \aap, 325, L17
\reference{Mar96} Markevitch, M., Mushotzky, R. F., Inoue, H.,
Yamashita, K., Furuzawa, A., \& Tawara, Y. 1996, ApJ, 456, 437
\reference{Or97} Orr, A., Molendi, S., Fiore, F., Grandi, P., Parmar, A.N., 
		\& Owens, A.
		1997, \aap, 324, L77
\reference{Ta94} Tanaka, Y, Inoue, H., \& Holt, S.S. 
		1994, \pasj, 46, L37 
\reference{Tuck1998} Tucker, W., \etal 1998, \apj, 496, L5 (T98) 
		

\end{references}
\end{document}